\documentclass[12pt]{article}
\usepackage{mathtools,amssymb,amsthm}
\usepackage{graphicx}
\usepackage{hyperref}
\usepackage{verbatim}
\usepackage{color}
\usepackage{float}
\usepackage{subcaption}
\usepackage[section]{placeins}
\usepackage{authblk}
\usepackage{amssymb}
\edef\restoreparindent{\parindent=\the\parindent\relax}
\usepackage{parskip}
\restoreparindent
\usepackage[normalem]{ulem}
\usepackage{fullpage}
 \def\pprec{\mathrel{\scalebox{.9}[1]{$\prec$}\mkern-4.5mu%
   \scalebox{.4}[1]{$\prec$}\mkern-4.5mu\scalebox{.4}[1]{$\prec$}}}

\graphicspath{{images/}}
\newtheorem*{proposition*}{Proposition:}
\newtheorem*{conjecture*}{Conjecture:}
\newcommand{\mink}{\mathbb M}
\newcommand{\cN}{\mathcal N}
\newcommand{\cA}{\mathcal A}
\newcommand{\link}{\prec \! \! * \, }

\newcommand{\dhor}{\twoheadrightarrow}

\newcommand{\av}[1]{\langle {#1} \rangle}

\newcommand{\fut}{\mathrm{Future}}
\newcommand{\past}{\mathrm{Past}}
\begin{document}

\title{ Null Geodesics from Ladder Molecules}
\author[1]{Anish Bhattacharya} 
\author[2]{Abhishek Mathur} 
\author[3]{Sumati Surya} 

\affil[1]{\small Indian Institute of Science, Bangalore, 560012, India}
\affil[2]{\small Chennai Mathematical Institute,  Kelambakkam, Tamil Nadu, 603103, India }
\affil[3]{\small Raman Research Institute, Sadashivanagar, Bangalore 560080, India}

\date{}
\maketitle
\begin{abstract}
 We propose a discrete  analogue of  null geodesics  in causal sets that are approximated by $\mink^2$, in the spirit of
 Kronheimer and  Penrose's ``grids'' and ``beams'' for an abstract causal space.  The causal set analogues are ``ladder molecules'', whose rungs are linked pairs of elements corresponding loosely to Barton et al's horizon 
 bi-atoms \cite{hormolone}.  In $\mink^2$ a ladder molecule traps a ribbon of null geodesics corresponding to a  thickened or fuzzed out horizon. The existence of a ladder between linked pairs of elements in turn provides a generalisation
 of  the horismotic relation to causal sets. Simulations of causal sets approximated by a region of $\mink^2$
show that ladder molecules are fairly  dense in the causal set, and provide a light-cone like grid.  
Moreover, similar to the uniqueness of null geodesics between horismotically related events  in $\mink^2$, in such causal sets there is a unique ladder molecule 
between any two linked pairs which are related by the generalised horismotic relation. 
 
  \end{abstract}

\vskip 0.75cm
  \begin{quote} 
{\sl ``To admit structures which can be very different from a manifold. The possibility arises, for example, of a locally
countable or discrete event-space equipped with causal relations macroscopically similar to those of a space-time
continuum.''} 

\hfill {\small{-- Kronheimer and Penrose on the aims of studying axiomatic causal spaces \cite{kp}.}} 
\end{quote}
\vskip 0.75cm

Kronheimer and  Penrose's (KP) abstract causal  spaces come equipped with the order  relations,  {$\preceq$} 
(causal) and $\pprec$ (chronological)  \cite{kp}\footnote{{Here  $\preceq$ is reflexive, i.e., $x \preceq x $  and
    $\pprec$ is irreflexive, $x \not {\! \!\!\! \pprec}  x $  }.}. These  are required to be acyclic (thus preventing closed causal and chronological
curves), with  each satisfying  transitivity individually, and together,  a 
mixed transitivity condition.  The KP causal spaces are therefore 
posets with respect to both  {$\preceq$}  and $ \pprec$ separately, with $\preceq$ strictly larger than the relation $\pprec$.  A third, 
the  horismotic relation $\rightarrow$,  is obtained from the first two by the simple exclusion: 
$x \preceq y, x \not\pprec y \Rightarrow x \rightarrow y$.  It is used by KP to build  ``girders'' and
``beams'' which provide the abstract analogs of the null-geodesics of a continuum spacetime.  

Causal posets are also the fundamental building blocks in  causal set theory \cite{blms,lr}. In this approach to quantum
gravity,  continuum spacetime is approximated by a locally finite poset or causal 
set, with the order relation $\preceq$ corresponding to the causal order of the spacetime. The requirement of local
finiteness allows a correspondence between the number of elements in a causal interval $n $ and the continuum spacetime volume $V$.  In the
continuum approximation which we denote by $C \sim (M,g)$  this is realised on average  as $\av{n} = \rho V$,   where $\rho^{-1}$ is the
spacetime discreteness scale.  The average is obtained  from a random Poisson sprinkling of points into $(M,g)$ at
density $\rho$,  with   probability $P_V(n) = \frac{ (\rho V)^n}{n!} e^{-\rho
    V}$. For each realisation,  a causal set is assembled from  the sprinkled points
using the induced 
causal ordering from  $(M,g)$.

The Poisson sprinkling process is 
uniform with respect to the spacetime volume which means that  null related elements are a set of measure zero. Thus the
elements  in $C \sim (M,g)$ are almost surely only related  via the 
chronological relationship $\pprec$ in $(M,g)$. {Because the chronological relation is irreflexive, it is therefore
  more appropriate to use the irreflexive $\prec$ symbol for causal sets.} The closest  analogue of 
a null-like relation between elements in a causal set is a {\sl link} or nearest neighbour relation $\link$ where   
$e \link e' \Rightarrow \, \not\!\! \exists \,  e'', \, e \prec e'' \prec e '$.    
For a causal set  $C$ which is approximated by $d$-dimensional Minkowski spacetime $ \mink^d$, for example, the elements linked to any $e \in C$ ``hug'' the
future and past light cone of $e$. However, these  links are
 almost surely time-like. While they can be used to roughly characterise the light-cones, they cannot substitute
for a null-relation; a  link in $C\sim \mink^d$  that hugs the light-cone in one frame can be far from null in another
frame. Thus one cannot
make a Lorentz invariant statement about the ``null-ness'' of a single link.  In order to obtain a truly  null-relation,
one 
would therefore need the right combination of  time-like and space-like related elements.

Null hypersurfaces are of course important in several contexts,  not least as black hole horizons. In
causal set theory there has been considerable success in extracting geometric information from the causal order,  including  discrete
versions of proper time, spatial topology, scalar field propagators and the Einstein-Hilbert action, which help in 
establishing a robust correlation between the continuum approximation and the underlying causal set ensemble. However,
an intrinsic definition of null hypersurfaces is still lacking. {In \cite{hormolone}  ``horizon
molecules'',  which are a class of  sub-causal sets,  were defined using the help of a continuum spatial hypersurface $\Sigma$ and a
null hypersurface $\cN$. $\Sigma$ and $\cN$ divide the spacetime into four regions $M_-^- \equiv J^-(\Sigma) \cap
J^-(\cN)$,  $M_+^- \equiv J^-(\Sigma) \cap J^+(\cN)$,  $M_-^+ \equiv J^+(\Sigma) \cap J^-(\cN)$ and  $M_+^+ \equiv J^+(\Sigma) \cap
J^+(\cN)$. Labelling the corresponding causal set regions $C_-^-, C_+^-, C_-^+, C_+^+$, respectively, the horizon
$n$-molecule is a subcauset $\{p, q^{(1)}, ...,q^{(n)}\}$ such that (i) $p \prec q^{(k)}$  $\, \forall k = 1,2,...n$, (ii)
$p \in C_-^-$, (iii) $q^{(k)} \in C_+^- \, \, \forall k = 1,2,...n$ (iv) $\{ q^{(1)}, ...,q^{(n)}\}$ are the only elements in
both $C_+^-$ and in  the future of $p$. } 
It was shown that in the limit of large sprinkling density $\rho$, the number of horizon
molecules is proportional to the area of the intersection $\cA = \Sigma\cap \cN$.  

While this construction is very powerful and works for any pair $(\Sigma, \cN)$,  the question remains -- what
intrinsically is a horizon in a causal set?  Inspired by the KP construction of null geodesics as
``beams'' as well as the 
horizon molecule construction we take a first step in  this direction.  Ours is a proposal for constructing
null geodesics in $C \sim \mink^2$; since   each null geodesic in $\mink^2$ is also a null hypersurface trivially, it
is also therefore a proposal for a causal set horizon.  Conversely,  it allows us to define a discrete horismotic
relation. We now  describe this proposal and show results from simulations that lend support to
it. Our discussion is strictly for $d=2$;  the analogous construction in  $\mink^d $ for $ d >2$ does not result from a
straightforward generalisation of the $d=2$ case.  At the end of this paper we discuss a possible generalisation to
higher $d$.

\begin{figure}[h]
 \centering 
   \includegraphics[scale=0.2]{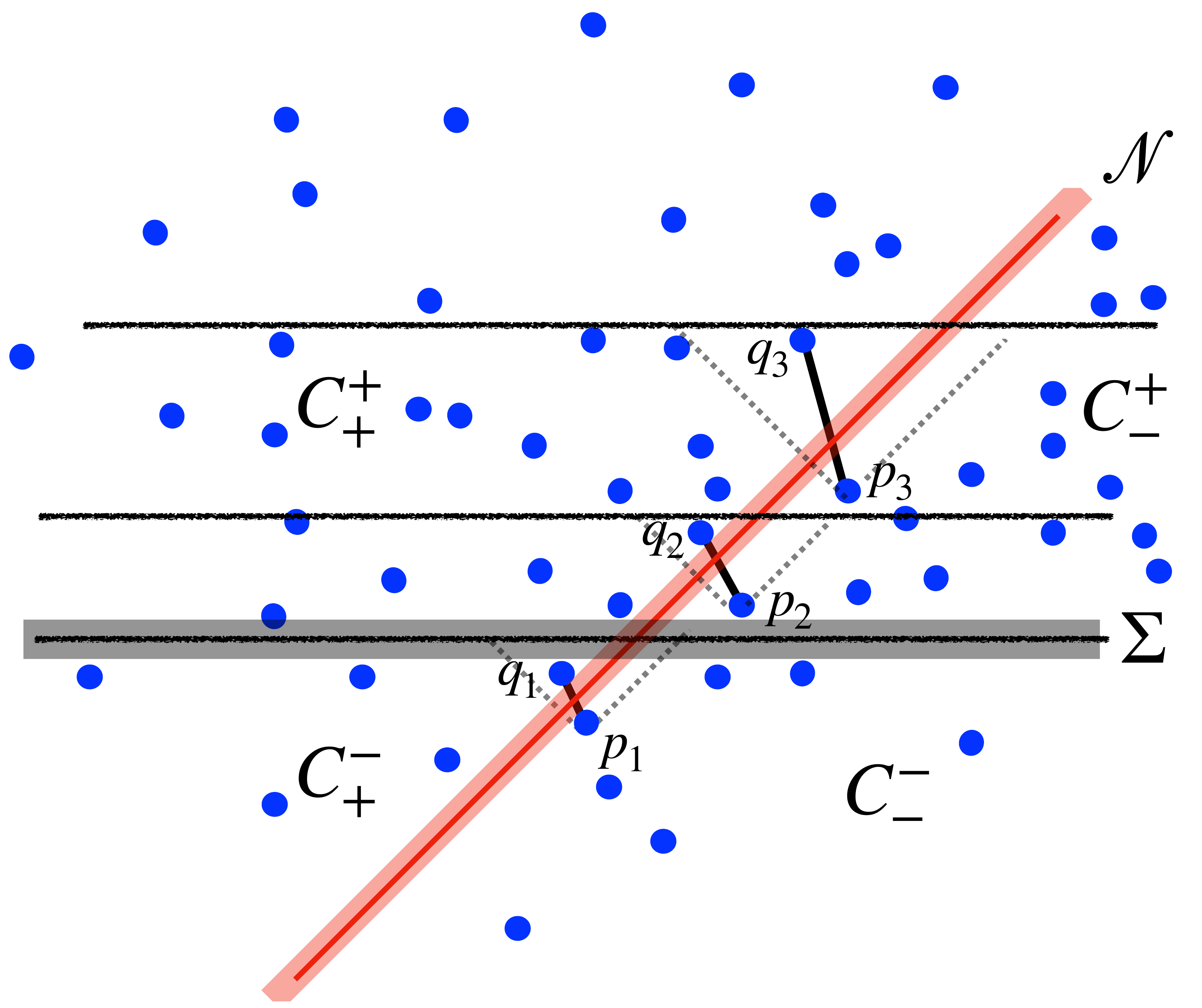}
   \caption{The horizon bi-atoms  $H_2^i$ associated with a family of spatial hypersurfaces which intersect a null
     hypersurface $\cN$. These are  ``stacked'' one on top of the other but do not form a ladder molecule. $\cN$ and the
     bottom-most spatial hypersurface $\Sigma$ are also ``thickened'' to depict a family of $(\cN, \Sigma)$ which share
     the same horizon bi-atom $H_2^1$ made up of the linked pair $ \{p_1,q_1 \}$. The regions $C_{\pm}^\pm\subset C$ depicted are
     with respect to $\Sigma$.  
   }
  \label{fam.fig} 
\end{figure}

We begin with the definition of the simplest  horizon molecule or bi-atom $H_2$ associated with $(\mathcal N,
\Sigma)$. This is a linked pair $(p,q)$ in $C \sim (M,g)$ where (a) $p \in I^-(\mathcal N) \cap I^-(\Sigma)$,(b) $q \in
I^+(\mathcal N) \cap I^-(\Sigma)$, (c) $q$ is the only point in $I^+(p) \cap I^-(\Sigma)$. This implies that $q$ is maximal
in  $C|_{I^+(\mathcal N) \cap I^-(\Sigma)}$ and  $p$ is
maximal in $C|_{ I^-(\mathcal N) \cap I^-(\Sigma)}$. 
Because of the fundamental discreteness of causal sets  $H_2$ is also a horizon
molecule for an entire  continuum family of $\{ (\cN, \Sigma)\}$. Fig \ref{fam.fig}  depicts the associated
``thickened''  horizon and spatial hypersurface for a single $H_2$. This suggests that any discrete analogue of
a horizon  will correspond at  best  to a thickened horizon in the continuum. 

In order to motivate our construction we  consider a discrete family of $\Sigma_i$ and the  
associated horizon molecules $H_2^i$ for each $\cN \cap \Sigma_i$ as shown in Fig \ref{fam.fig}.  Given such a ``stack''
of $H_2^i$, we can think of  them as guides for a thickened horizon. However since the $H_2^i$ are not intrinsically
defined, this is not sufficient for our purpose.   Instead we consider  stacks of  
linked pairs $h_i=(p_i, q_i), \,  p_i \link q_i$. In order to recover a null geodesic, we must  stack these one on top
of the other in a suitable and intrinsic manner so that, when embedded in the continuum they trap a ribbon of null
geodesics.  We denote the causal interval by  $[p,q] \equiv \{r \in C|  r \in \fut(p) \cap \past(q)
\}$, where $\fut(p) \equiv \{r \in C | p \prec r \}$ and  $\past(p) \equiv \{r \in C | r \prec p \}$.    Importantly,
because of the use of the irreflexive relation $\prec$,  $p,q \not\in [p,q]$ 

We construct a ladder molecule $L_k$ step by step as follows. Let $L_1 \equiv  h_1$
where $h_1=(p_1,q_1), \, p_1 \link q_1$ be the first rung.  The next rung  $h_2  = (p_2,q_2), p_2 \link q_2$  must 
satisfy the stacking condition 
\begin{equation}
p_1 \link p_2, q_1 \link q_2, p_1 \prec q_2, |[p_1,q_2]|=2. 
\end{equation}
 The rung $h_2$  is thus ``stacked'' on top of the rung $h_1$  in a causally rigid manner since no elements other than those in
 $h_1,h_2$ are allowed in the interval $[p_1, q_2]$.  $L_2$ is then the  
4-element diamond sub-causal set, with the pair $q_1,p_2$
space-like to each other. {More generally, we  build  $L_k$ from $L_{k-1}$ by stacking $h_k=(p_k,q_k)$ on top of $L_{k-1}$
such that
\begin{equation}
  p_{k-1} \link p_k, q_{k-1}\link 
  q_k, |[p_{k-i},q_k]| =2i, |[p_i,p_j]|=j-i-1, |[q_i,q_j]|=j-i-1. 
\end{equation}
Thus, $|[p_1,q_k ]|=2(k-1)$, with the pairs $p_i, q_{j}$ being  space-like to each other for all $j<i$.} This 
volume rigidity that goes all the way down the ladder  ensures that it is ``as straight as possible''. It is important to note  here that although inspired by the horizon molecules, typically 
stacks of horizon bi-atoms do not form ladders as depicted  in Fig \ref{fam.fig}.    

\begin{figure}[h]
 \centering 
  \includegraphics[scale=0.5]{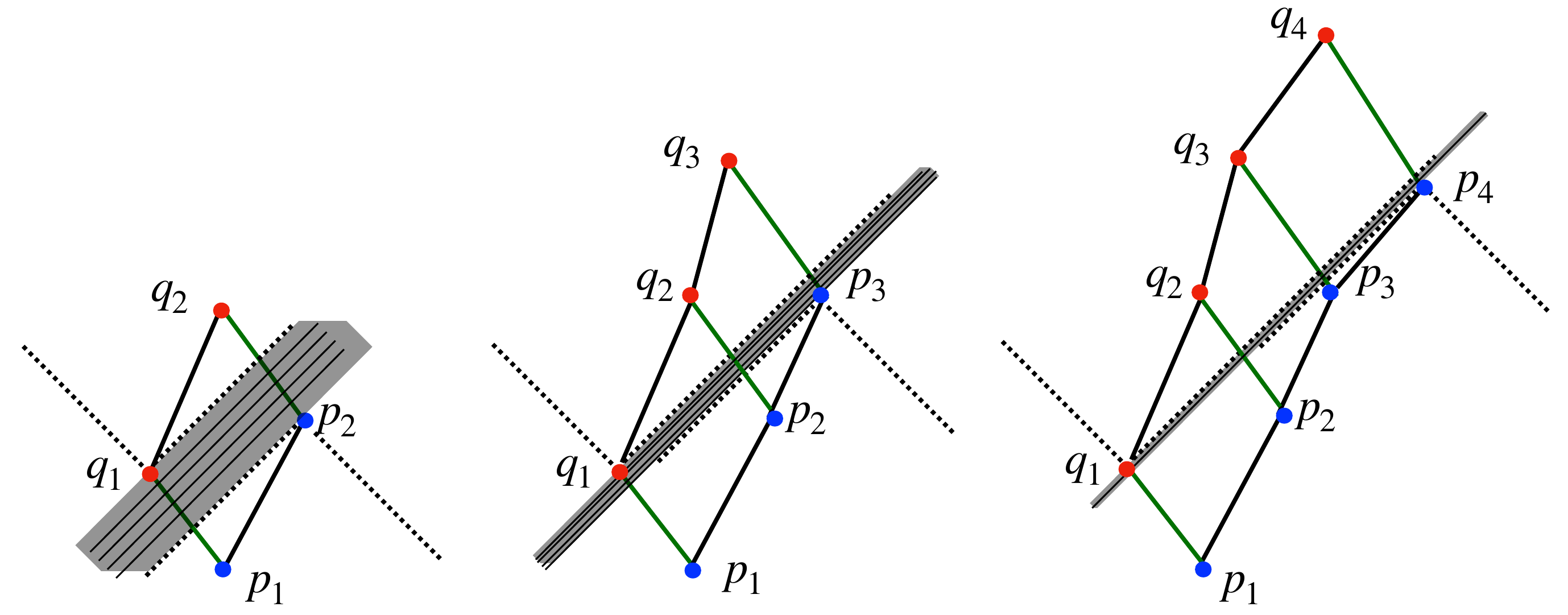}
  \caption{Illustrations of ladder molecules $L_2, L_3, L_4$ and the associated ribbons of null geodesics. As more rungs
  are added the  ribbons become narrower. }  
  \label{l2l3l4.fig} 
\end{figure}

As shown in Fig \ref{l2l3l4.fig} the  $L_k$ trap a ribbon of null geodesics  in $\mink^2$, which we define as follows.  Let $\mathcal P$ denote the set of all
inextendible null geodesics  in $\mink^2$. The ribbon of null geodesics associated with an $L_k$ is   
\begin{equation}
\cN_k(L_k)\equiv  \{ \eta  \in \mathcal P|  \eta \cap J^+(q_1) = \emptyset, \eta  \cap J^-(p_k) = \emptyset\}. 
  \end{equation} 

We now use  the ladder molecules  to generalise  the horismotic relation to causal sets. We say that the linked pairs $h=(p,
q), p \link q $ and   $h'=(p',
q'), p' \link q'$ are discrete horismotically related  
\begin{equation}
h \dhor h' \, \, \mathrm{if} \, \, \exists \, \, L_k, \, k \geq 2,  \, \, L_k|_1=h, L_k|_k=h'.  
\end{equation}
where $L_k|_i = h_i$ denotes  the  $i$th rung of an  $L_k$. As in the case of the  horismotic relation,  $\dhor$ does not satisfy
 transitivity: 
 \begin{equation}
h \dhor h', h' \dhor h'' \not\Rightarrow  h \dhor h'', 
   \end{equation} 
   since the two ladders $L$ from $h$ to $h'$ and $L'$ from $h'$ to $h"$ while sharing a rung $h'$ (and thus being
   causally ordered) may not
   satisfy the various rigidity conditions between elements in $L$ and those in $L'$.  This is similar to  the
   horismotic relation in the continuum where $a \rightarrow b$, $b \rightarrow c \, \, \not \!\Rightarrow \,  a \rightarrow
   c$.  We know
   that unlike the chronological relation, if $p \rightarrow q$ then there exists  a unique future directed null
   geodesic from $p$  to $q$ in $\mink^2$. Is this also true for the relation $\dhor$?  As we discuss below there is numerical
   evidence to suggest  that this is indeed the case.

\begin{figure}[h]
 \centering 
  \includegraphics[scale=0.5]{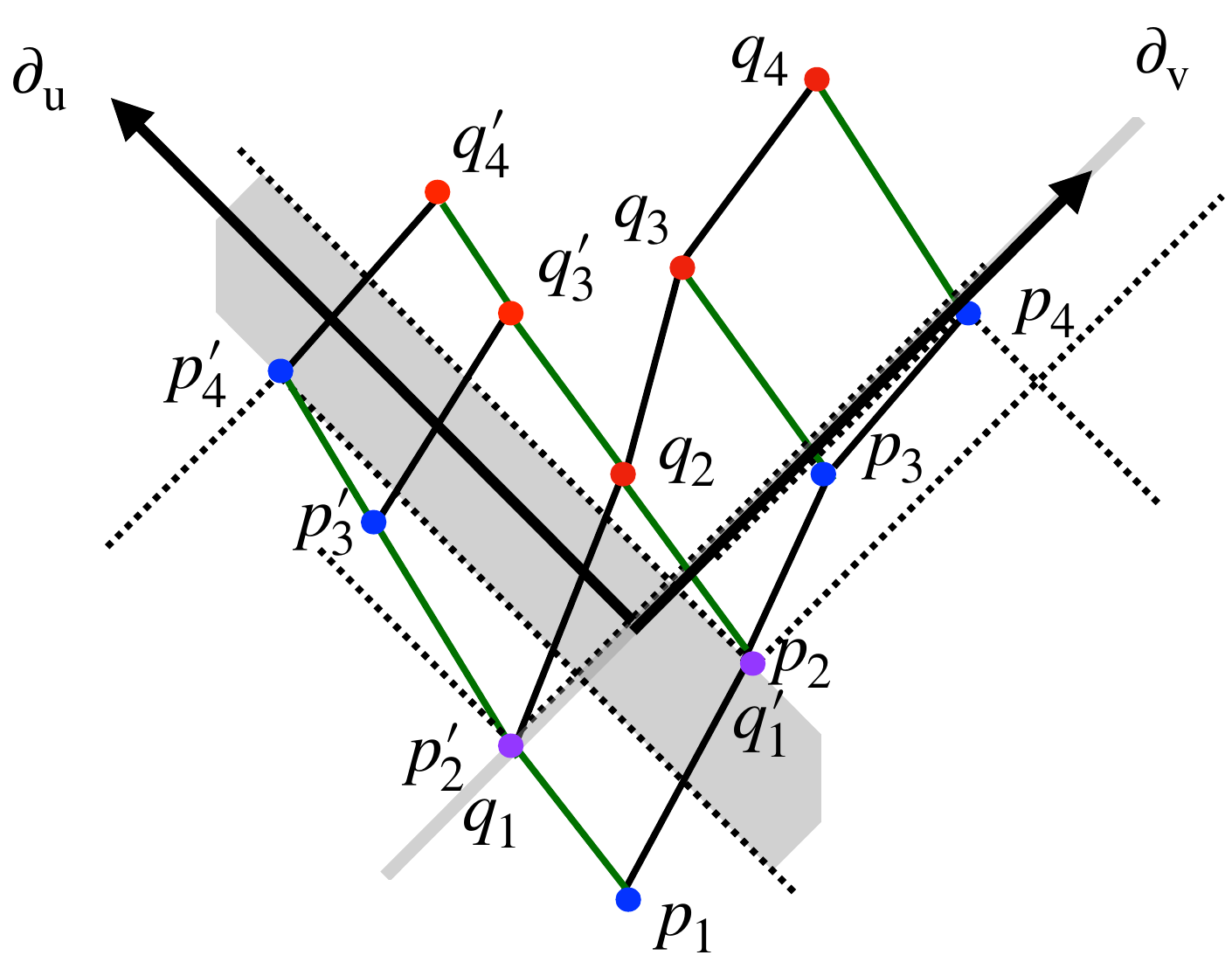}
  \caption{An illustration  of a $V_4$ sub-causal set and the directions $\partial_u,\partial_v$ that can be obtained
    from it.}  
  \label{uv.fig} 
 \end{figure}

In $\mink^2$ we can go further since we need only two independent null directions $\partial_u, \partial_v$ to define a
tangent basis. As we have defined
them, the ladders do not carry an intrinsic direction that would help assign a family of  right movers or  left movers to them
without the embedding.  Consider  the causal diamond $L_2$ with rungs $h_1,h_2$, $h_1 \dhor h_2$.  Since the pair
$q_1,p_2$ are space-like, with both linked to $p_1$ to the past and $q_2$  to their future it can also be viewed as the
ladder $L_2' $ with $h_1'=(p_1,p_2)$, $h_2 '= (q_1, q_2)$, where $q_1 \leftrightarrow p_2$. Thus, one
can ``grow'' a ladder $L_k, L_k'$ either using  the $h_2$  rung or the  $h_2'$ rung as shown in Fig \ref{uv.fig}.  These
two ladders then correspond to the two null directions $\partial_u, \partial_v$  ``at'' $L_2$.  We will call this double
ladder a $V_k$ sub-causal set.  

Since we  were motivated by the horizon bi-atoms, we can now ask whether the  rungs of the ladder molecules are in fact
horizon molecules of a given spatial hypersurface. In the continuum this is trivially the case since we can always
``fit'' in a  $\Sigma$.  In the discrete setting the  analogue of a spatial hypersurface is an antichain. If we take the antichain $\mathcal A_i$ associated with $h_i=(p_i,q_i)$ as one
that  ``threads''  through $q_i$, but is to the past of $p_{i+1}$, then the intersection of the discrete horizon with
$\mathcal A_i$ is just $q_i$. The ``area''  of this intersection is then proportional to the number of horizon molecules as in \cite{hormolone}.  
 
The first question we must address is whether ladders occur often enough in $C \sim \mink^2$ to make this definition
a useful one in our characterisation of null geodesics in $C$. Starting with a linked pair $h_1=(p_1,q_1)$, how likely
it is that there is a  next rung $h_2=(p_2,q_2)$, and a next one and so on? Since a ladder $L_k$ embeds\footnote{An
  embedding $C \hookrightarrow M$ does not require the number-volume correspondence to hold. Not 
 all causal sets embed in $\mink^2$. However,  the fact that the ladder {does}  embed {\it does not} mean that it is in any sense
  typical.} in
$\mink^2$ for any  $ k > 1 $, an  $L_k$ with $L_k|_1=h_1$ almost surely exists  in $\mink^2$ since there is infinite
room,  so to speak,  to find the next
rung and the next\footnote{Spaces of infinite  extent allow all kinds of possibilities for causal sets, but may not be
  physically relevant. For example,
  while  a void the size of the present universe almost surely occurs  in
  $\mink^d$, the probability for a Fermi-radius sized  void within the Hubble radius 
is essentially zero, as shown in   \cite{swerves}.}.
What is more relevant however is the density of ladders in a given bounded region of $\mink^2$.  In  \cite{intervals}
the abundance of $k$-element intervals in a causal diamond in $\mink^d$ was calculated analytically as  a function of
sprinkling density and the volume of the causal diamond. While the  ladders are $2(k-1)$ element  intervals, they are
not the only causal sets in this class. For example, for $k=2$, there are two causal sets which are $2$-element
intervals: the causal diamond poset $L_2$ and the $4$-element totally ordered set or chain. The calculations of
\cite{intervals} do not tell us anything about the more detailed distributions of specific sub-causal sets.

Since we have defined the $L_k$ using the the discrete intervals the likelihood can be assessed by looking at the
associated continuum volumes in a Poisson sprinkling.  For every  $x \in \mink^2$
there are  invariant hyperbolae defined by the volume $V$ of the Alexandrov interval between any point on this hyperbola and
$x$ (or equivalently, its proper time $\sqrt{2 V}$ to $x$)  both to the future and the past.  Let  $v_k(x)$ denote 
the region between the future invariant hyperbolae of discrete volume $k - \sqrt{k} $ and  $k + \sqrt{k}$ about  $x$
where $k=\rho V$, and we have included the Poisson fluctuations of order $\sqrt{\rho V}$. 

Let us start with the likelihood of finding an $L_2$ associated with a given $p_1$.  For $p_1\link q_1$ the
interval $[p_1, q_1]$ should be empty, and hence  $q_1$ is most likely to lie in the region $v_1(p_1)$ between the future light
cone of $p_1$, i.e., at $k=1-\sqrt{1} =0$  and the invariant hyperbola at 
discrete spacetime volume  at $k=1+\sqrt{1}=2$.  This is a region of infinite spacetime volume and  hence 
almost surely  there exists a pair $h_1=\{p_1, q_1\}$. Next, we look for $p_2$ where $p_1 \link p_2$. Since
the volume of $v_1(p_1)$ is infinite, such a $p_2$ can again almost surely be found. The conditions on $q_2$ are however much
more stringent since not only should $p_2 \link q_2$ and $q_1 \link q_2$, but also $|[p_1,q_2]|=2$, i.e.,  $q_2 \in v_1(q_1)
\cap v_1(p_2) \cap v_3(p_1)$.  For such a  $q_2$ to be sufficiently likely, the discrete volume $\rho \mathrm{vol}(v_1(q_1)
\cap v_1(p_2) \cap v_3(p_1) ) \gtrapprox  1$.    

\begin{figure}[h]
  \begin{center}
    \vspace{0.5cm}
 \includegraphics[height=5cm]{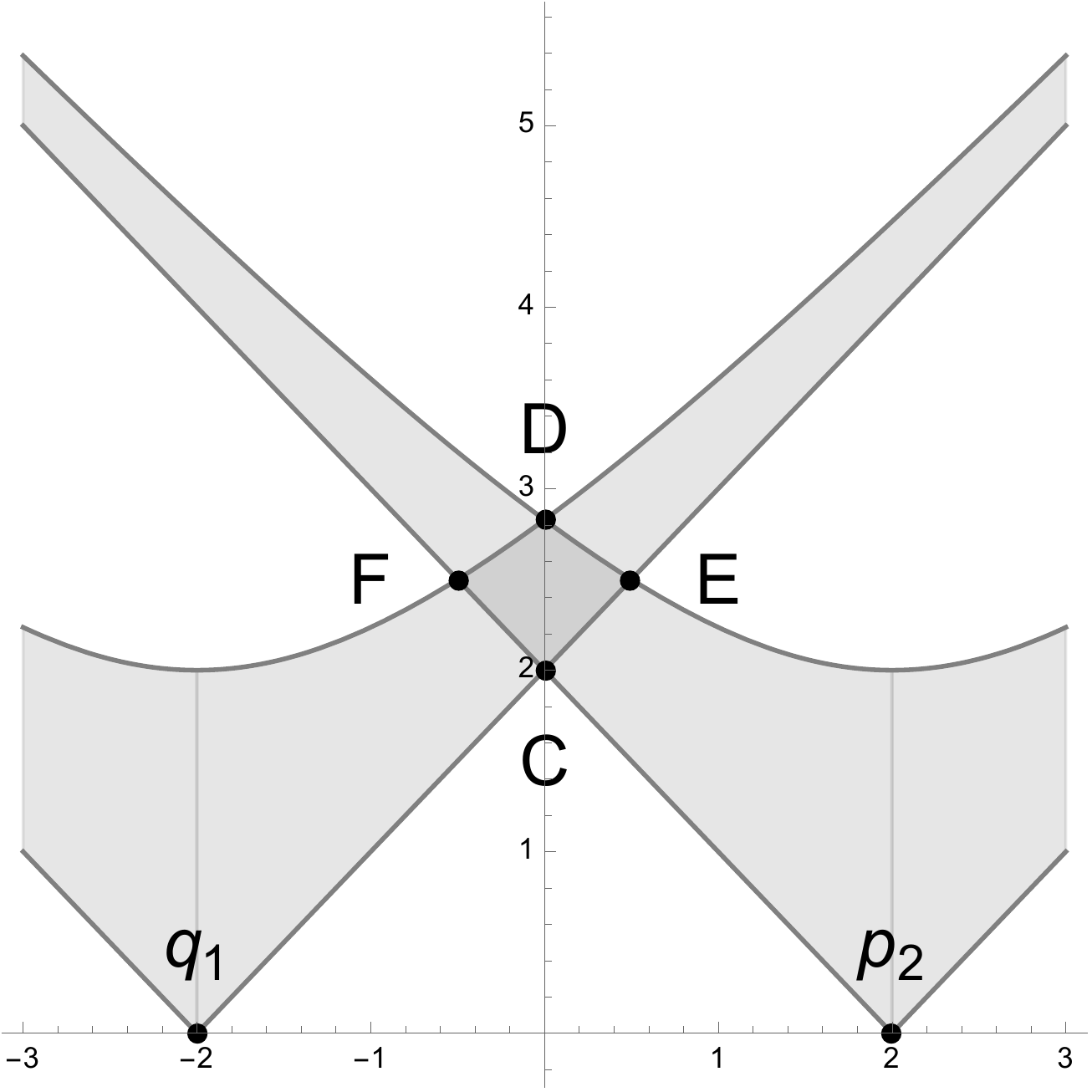} \hskip 2cm
 \includegraphics[height=5cm]{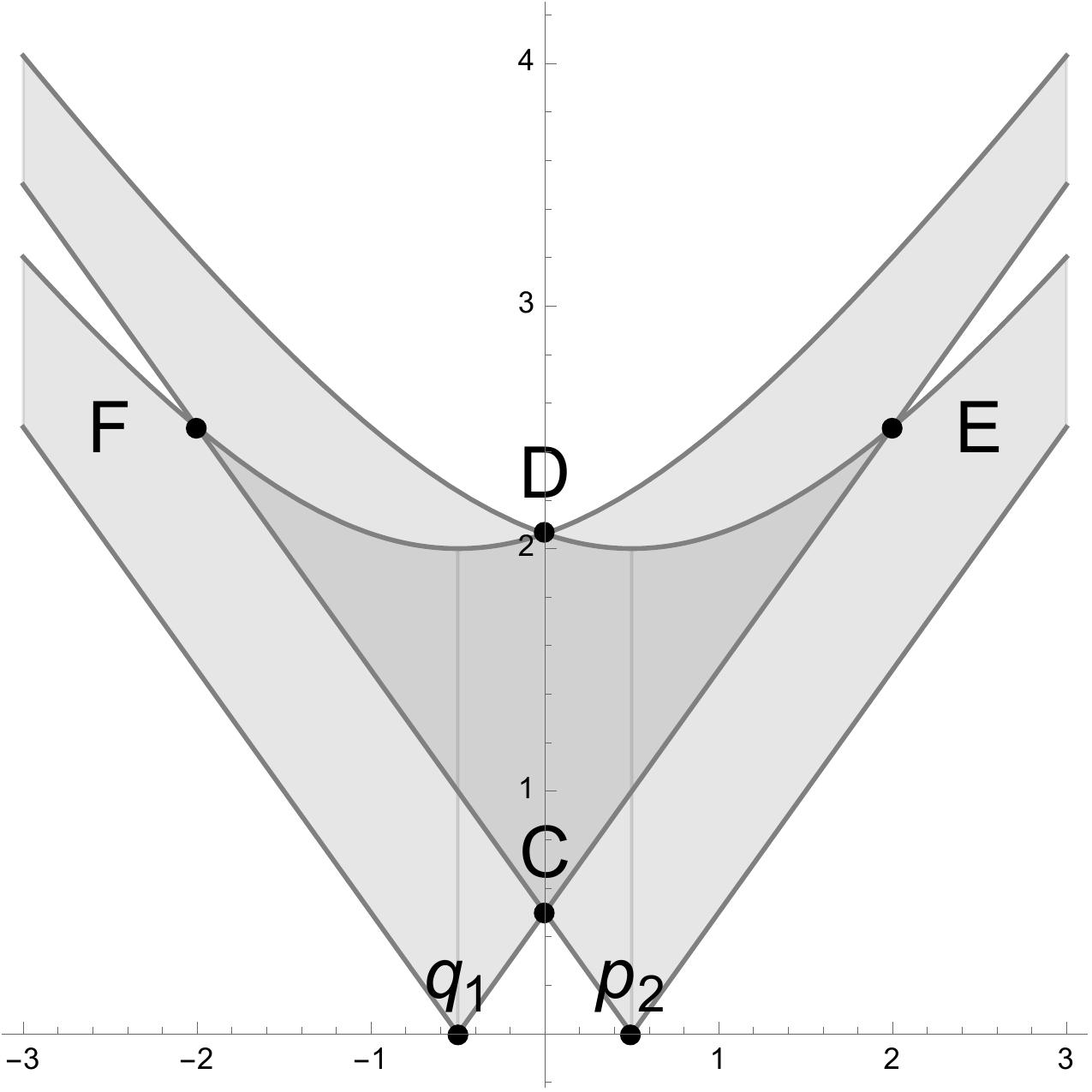} 
                \caption{As the proper distance between $q_1$ and $p_2$ decreases, the overlap region $CEDFC$ (shaded
                  dark gray)  increases monotonically. } \label{v1q1p2.fig}
                 \vspace{0.5cm}
\end{center} 
\end{figure}

First we look at the region $ CEDFC \equiv v_1(q_1) \cap v_1(p_2) $ which is  itself  finite as shown in Fig \ref{v1q1p2.fig}, when 
the proper space-like distance $d=d(q_1,p_2)>0$.  We can always choose coordinates such that $q_1=(0,-d/2)$ and $p_2 =
(0,d/2)$. In this case, $C=(d/2,0), D=(\sqrt{d^2/4+4},0), E=(2/d+d/2,2/d), F=(2/d+d/2, -2/d)$.  $CEDFC$ is symmetric
about  $x=0$ and  bounded by the curves $t^2-(x+d/2)^2=4$ (upper boundary of $v_1(q_1)$),  
$t^2-(x-d/2)^2=4$ (upper boundary of $v_2(p_2)$) and the null ray  $t=x+d/2$  from $q_1$ and the null ray $t=x-d/2$
from  $p_2$.  Thus, its  volume is 
\begin{eqnarray} 
  V(d) & = & 2\biggl( \int_0^{2/d} \biggl( \sqrt{\biggl(x-\frac{d}{2}\biggr)^2+4}-(x+d/2) d x \biggr)  \biggr) \nonumber \\
  &=&  4 \sinh^{-1}\biggl(\frac{4-d^2}{4d} \biggr)   + 4 \sinh^{-1}\biggl(\frac{d}{4} \biggr) -2 + \frac{d}{4}
      \biggl(\sqrt{d^2+16} -d \biggr)  
\end{eqnarray} 
 which monotonically decreases with $d$, becoming infinite as $d \rightarrow 0$. For $d< d_0 \sim 2.23$ the discrete
 volume $ \gtrapprox 1$. 

\begin{figure}[h]
  \begin{center}  \vspace{0.5cm}
\includegraphics[height=4cm]{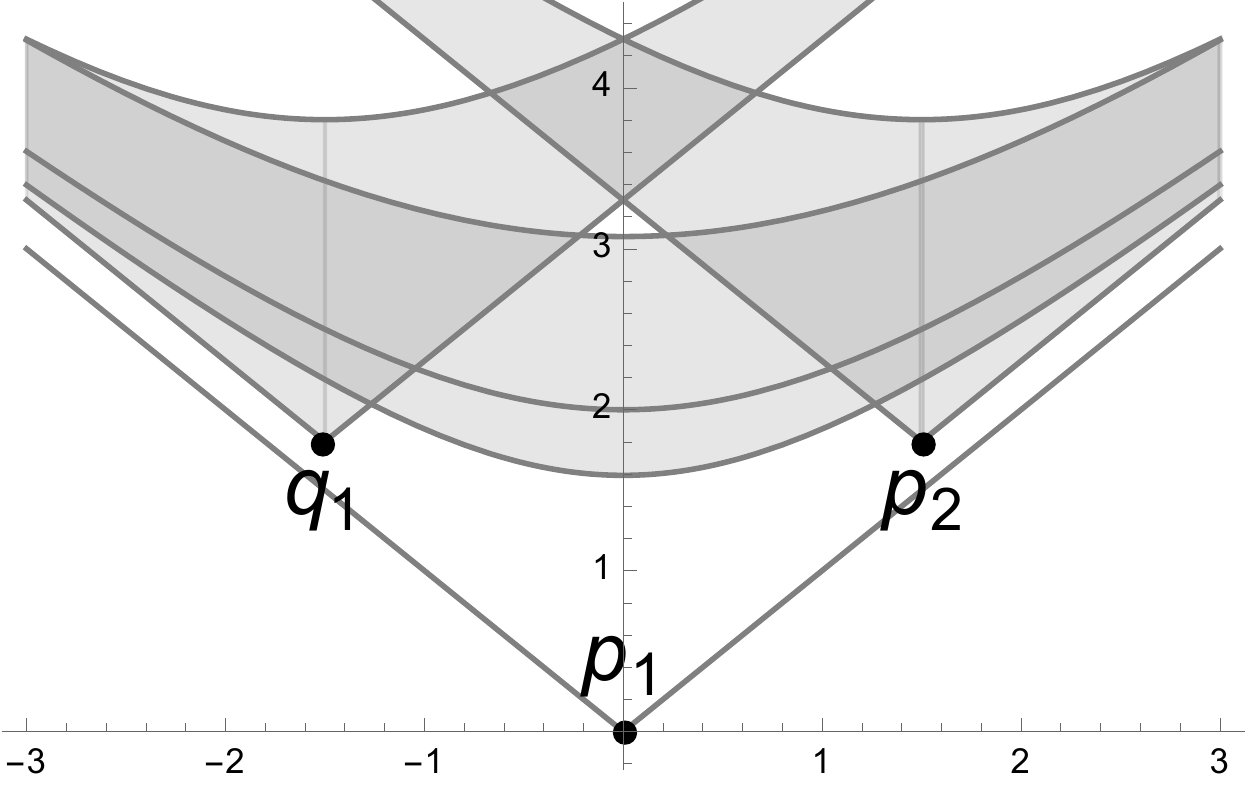} \hskip 2cm 
 \includegraphics[height=4cm]{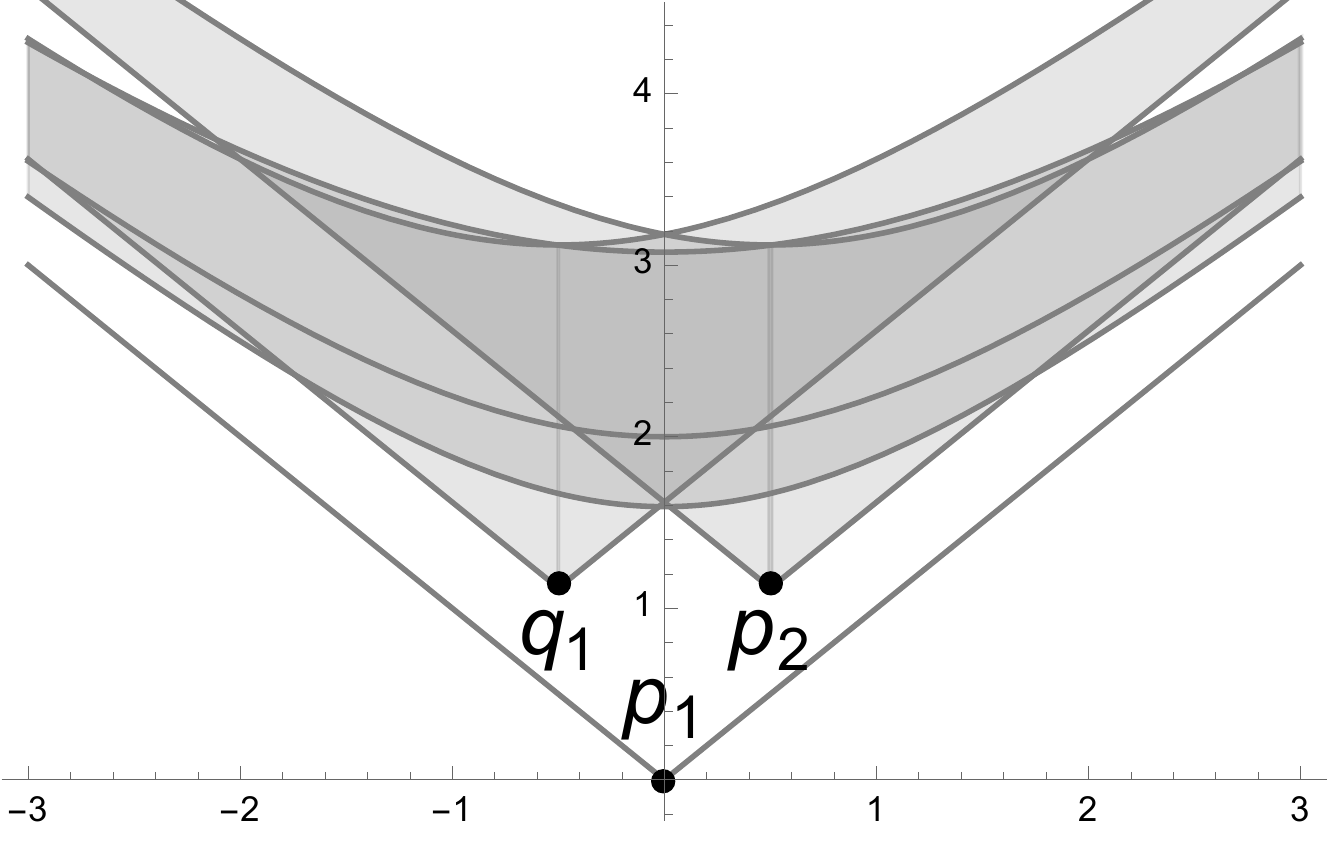} 
                \caption{The overlap  region $O\equiv v_1(q_1)\cap v_1(p_2) \cap v_3(p_1)$ for two different choices of $d$ and
                  $T$. $q_1, p_2$ lie in the the region $v_1(p_1)$ which  is unshaded, but overlaps with  the shaded
                  region $v_3(p_1)$.  In the figure on the left $d$ and $T$ are large enough that   $O$ is  empty: the
                  $CEDFC$ region lies entirely above $v_3(p_1)$.  In the 
                  figure on the right $d$ and $T$ values are small enough to make the overlap  (the darkest shaded region)  $O$  non-empty. } \label{hyp.fig}   \vspace{0.5cm}            
\end{center} 
              \end{figure}

The tricky part of course is to ensure that the further overlap with  $v_3(p_1)$ is large enough.   
Fig \ref{hyp.fig} illustrates the fact that when $d$ is too large the overlap is zero, while the overlap can be
significant for smaller $d$.  In order to calculate the volume of the overlap region, we make a 
simplifying assumption that $q_1$ and $p_2$ are symmetrically to the future of $p_1$, i.e., if $p_1=(0,0)$ then we assume  $q_1=(T,-d/2),
p_2=(T,d/2)$ (which is of course not strictly possible in the causal set.) The overlap region $CEDFC=v_1(q_1) \cap
v_1(p_2)$ has $C=(d/2+T,0), D=(\sqrt{d^2/4+4}+T,0)$ and $E=(d/2 + 2/d +T,2/d)$  and  $F=(d/2
+ 2/d +T,-2/d)$ while the boundary hyperbolae  of $v_3(p_1)$ intersect the $t$-axis at $\tau_-=(\sqrt{6-2 \sqrt{3}},0)$ and
$\tau_+=(\sqrt{6+2 \sqrt{3}},0)$. Here there are two parameters $d$ and $T$, but because the lower boundary of
$v_3(p_1)$ is at $3-\sqrt{3} < 2 $, the overlap with $CEDFC$ is not bounded from above as  $d \rightarrow 0$ and $T
\rightarrow 0$. For non-vanishing  values of $d$ and $T$ (which is almost surely the case for $\{p_1,q_1,p_2 \}$), the overlap is
finite, and the question is whether  there is a  ``sweet spot''  in $(d,T)$ for which this overlap is large but also
probable.  

\begin{figure}[h]
 \centering  \includegraphics[height=6cm]{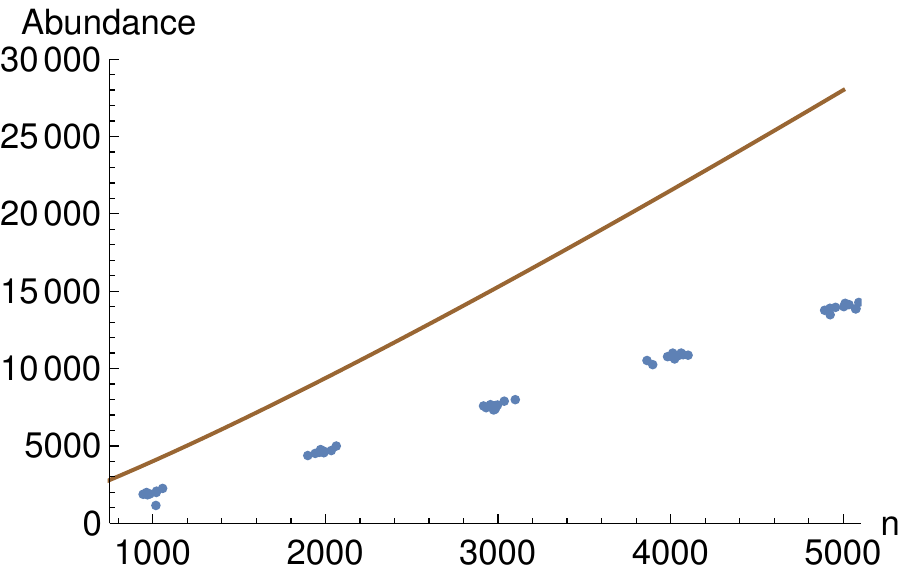} 
  \caption{The abundance of causal diamond intervals in a causal diamond in $\mink^2$ as a function of $n$. We  compare
    it with the analytic expression for the abundance of all $2$-element intervals depicted by a solid line.}  
  \label{abun-diamond.fig} 
 \end{figure}

Calculating this analytically is messy, since there are a large number of  possible types of intersections of $CEDFC$ with
$v_3(p_1)$ leading to different formulae for the volume.   We instead turn
to numerical simulations to  help us determine the abundance of the $L_2$ in a causal diamond. Fig
\ref{abun-diamond.fig} gives an idea of the growth of the  $L_2$  as a function of $n$.
Comparing with the analytic expression of \cite{intervals} we see that the diamonds are roughly the same in number as
the remaining $2$-element intervals, namely the $4$-element  
chain.  Note  that the $4$-element chains are intrinsically more ``time-like'', and hence can stray further
away from null directions. The  causal diamond interval on the other  hand  has
both  time-like ($p_1$ and $q_2$) and space-like ($q_1$ and $p_2$) relations, which force it into a more null-like
configuration. 

In the rest of the paper we present results  on  numerical simulations of causal sets. We find that the $L_k$ for
$k=2,3,4$ are not  rare even in finite regions of $\mink^2$. We calibrate the density of these discrete null geodesics
by using the embedding in $\mink^2$ to obtain the null ribbons and find that they form a dense null grid on the causal set. We also find that that for small $k$ values the $V_k$ causal sets while not
common, do exist, thus providing local  $\partial_u, \partial_v$  directions in the causal set.  Finally we find that the ladders satisfy a
uniqueness  condition, ``upto $q_k$''. This suggests a discrete version of Penrose's Proposition 2.19 \cite{penrose}
(see below) for
the generalised horismotic relation $\dhor$.  

\begin{figure}[H]
 \centering 
  \includegraphics[scale=0.5]{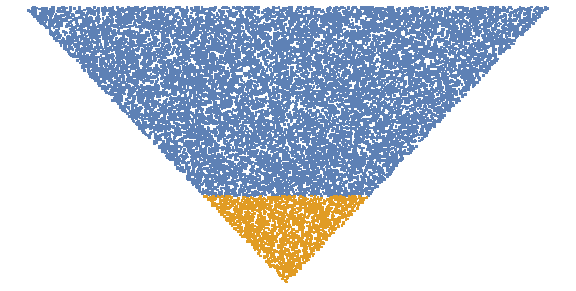} 
  \caption{The $p_1$ are chosen from the bottom $\sim n/10$th set of sprinkled elements.}  
  \label{choicep1.fig} 
 \end{figure}

For numerical convenience in what follows we  consider  sprinklings into  a half diamond in $\mink^2$ rather than the
full diamond, for
\begin{equation}
  {n \sim 500, 1000, 2000, 3000, 5000, 7000, 10000, 12000, 15000, 18000}.
\end{equation}
and 
perform $10$ trials for each value of $n$.  We search for the 
$L_k$  starting with a $p_1$ in the bottom $n/10$ of the elements as shown in  Fig \ref{choicep1.fig}.  We 
restrict our search to $k=2,3,4$, since already for $k=4$, the abundance is very small.  

\begin{figure}[H]
 \centering 
  \includegraphics[height=6cm]{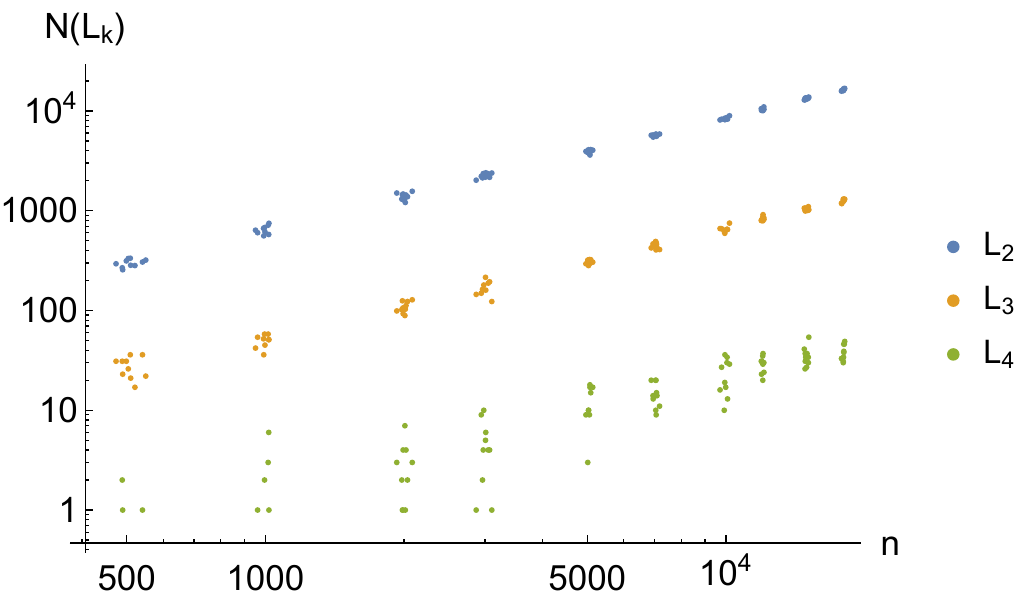}
  \caption{A log-log plot of the abundance  of $L_k$ with $n$.}  
  \label{abundance.fig} 
\end{figure}

 Fig \ref{abundance.fig} shows the number of ladders for  $k=2,3,4$ as a function of $n$ on a log-log plot. There is a
 clear scaling behaviour of   the abundance of $L_k$,  i.e.,  {$N(L_k) \sim a_k n^{\alpha_k}$, with $a_k$ a decreasing function of $k$ with the
   exponent $\alpha_k$ being  approximately 
 $k$ independent. Thus the abundance decreases with $k$ as expected, with a significant population of $k=2, 3$ ladders compared to $k=4$. }   
{The $L_k$ are  moreover statistically significant enough to form a dense  ``grid''  of null geodesics in $\mink^2$.
In Fig \ref{grid.fig} we plot  the null ribbons associated with every $L_3$ and $L_4$  which has been generated in  a particular
sprinkling with ${n \sim 18,000}$.  These  form a  dense null grid in $\mink^2$, the former more dense than the latter. 
\begin{figure}[H]
  \centering 
  \centerline{\begin{tabular}{ccc}
    \includegraphics[height=4cm]{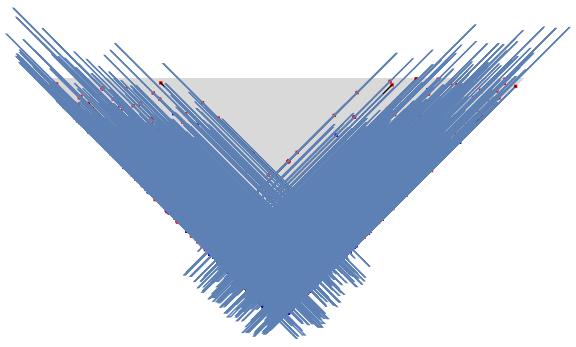} &   \hskip 1cm
                                                             \includegraphics[height=3cm]{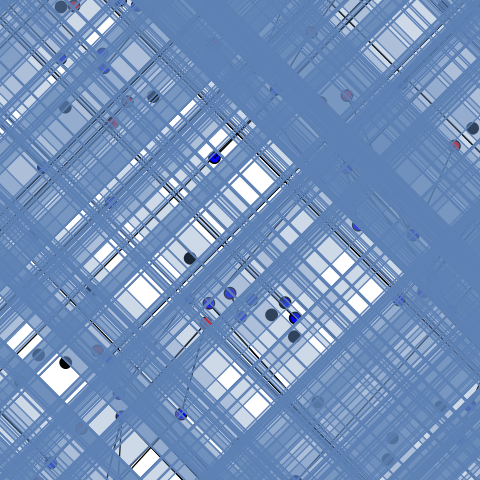}&  \hskip 1cm
                                                         \includegraphics[height=4cm]{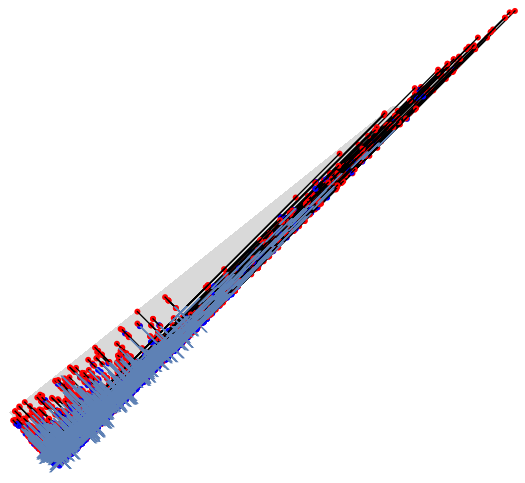} \\             
 \includegraphics[height=4cm]{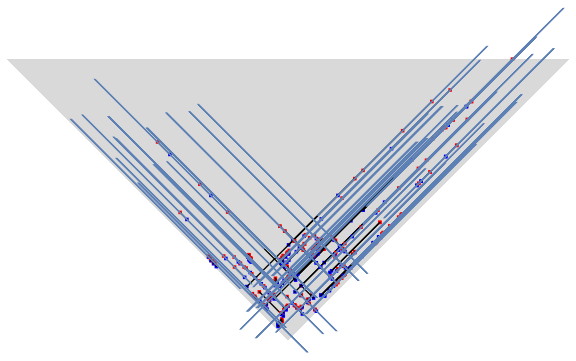} &  \hskip 1cm \includegraphics[height=3cm]{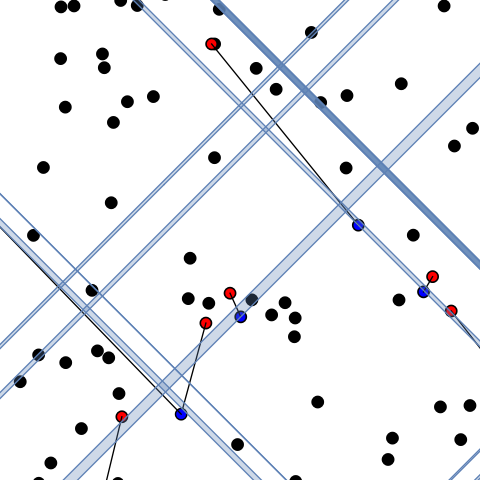}&\hskip 1cm
                                                         \includegraphics[height=4cm]{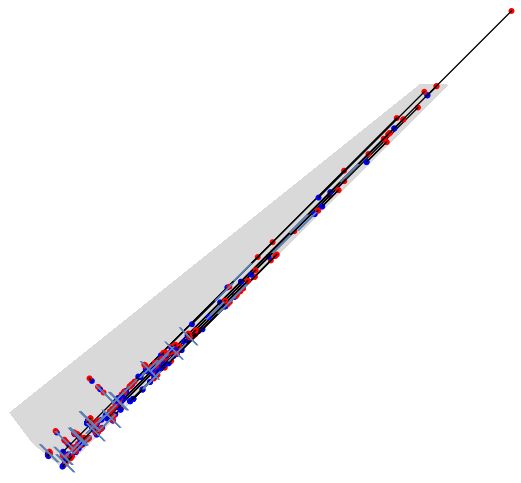} \\
                 \end{tabular}}
                \caption{The null grid formed by $L_3$ and $L_4$ ladders in an $n \sim 18, 000$ sprinkling. We show 
                  the same zoomed-in region in  both cases, as well as the entire Lorentz transformed region. The grid
                  for $L_3$ is far more dense than for $L_4$ and is 
                  fine enough to reach the discreteness scale.} \label{grid.fig}    
            \end{figure}
            
In Fig \ref{singleladder.fig}  we show  a specific example of an $L_4$ ladder and the associated null ribbon for different
boost parameters, for an {$n\sim 18,000$}  sprinkling.  As expected, the null  ribbon ``fattens'' or ``thins'' depending on the choice of $\beta$. In $\mink^2$ there is a simple Lorentz
invariant quantity 
associated with the ribbon. This is the causal volume associated  with the unique 
pair $(x,y)$, 
$x \rightarrow q_1, x \rightarrow p_n$ and $q_1 \rightarrow y, p_n \rightarrow y$. The volume of $[x,y]$  is therefore
an invariant  under a  Lorentz transformation. Thus, one can  squeeze  or fatten the ribbon without changing this
volume. 

\begin{figure}[H]
 \centering {\begin{tabular}{ccc}
               \includegraphics[scale=0.5]{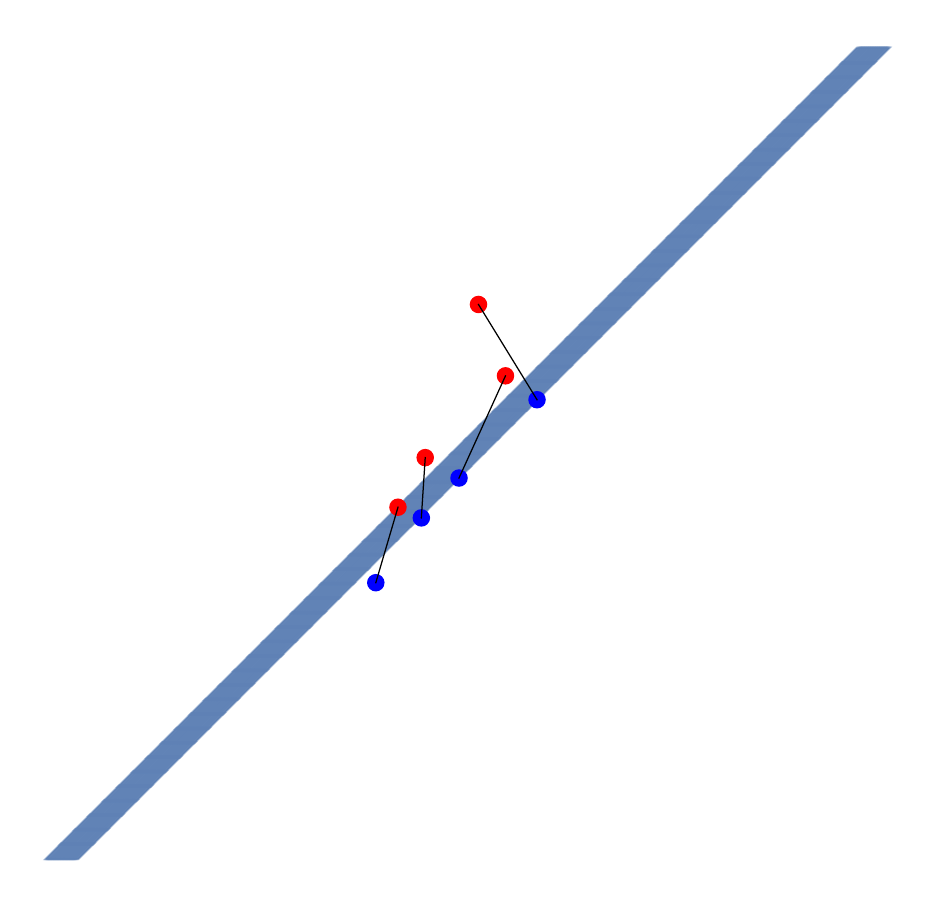}&  \includegraphics[scale=0.5]{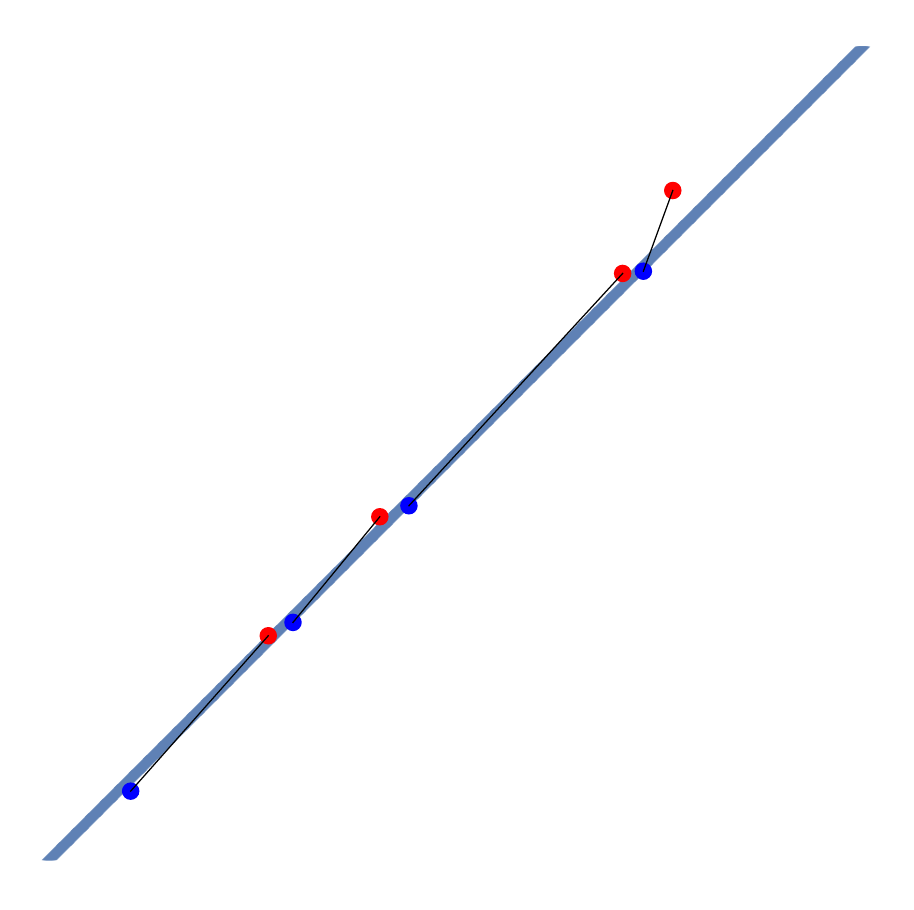} &
                                                                        \includegraphics[scale=0.5]{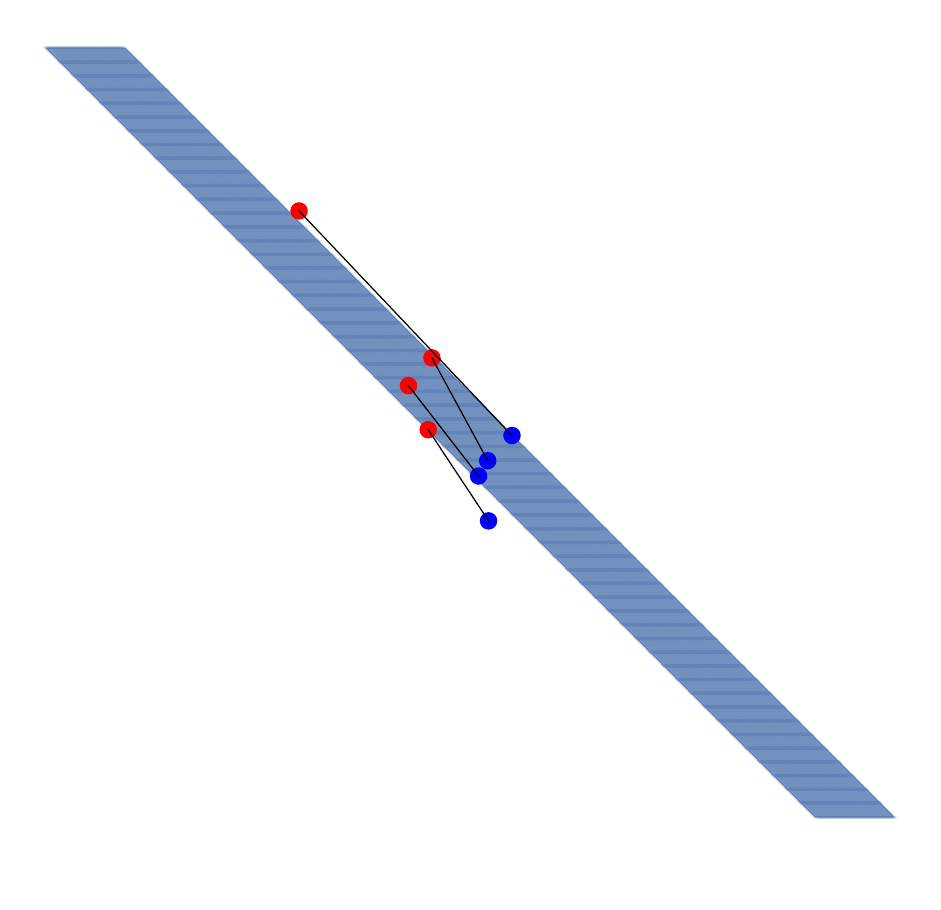} 
                                                                                                                                           \\ 
 \end{tabular}}               
  \caption{An $L_4$ ladder in an $n \sim 18,000$ sprinkling with boost parameters  (a) $\beta=0$ (b) $\beta=0.8$ (c) $\beta=-0.8$. The $p_i$ are
  depicted in blue while the $q_i$ are depicted in red.}  
  \label{singleladder.fig} 
\end{figure}

Our simulations also give rise to  intersecting or $V$-type ribbons of the type shown in Fig \ref{uv.fig}. These are
however  far fewer than the ladders. For $V_3$, we show  the number of occurence  in Fig \ref{v-fig.fig}. In Fig \ref{vex.fig} we show the only
example of a $V_4$ that resulted from our simulations.
\begin{figure}[H]
  \centering  \includegraphics[height=6cm]{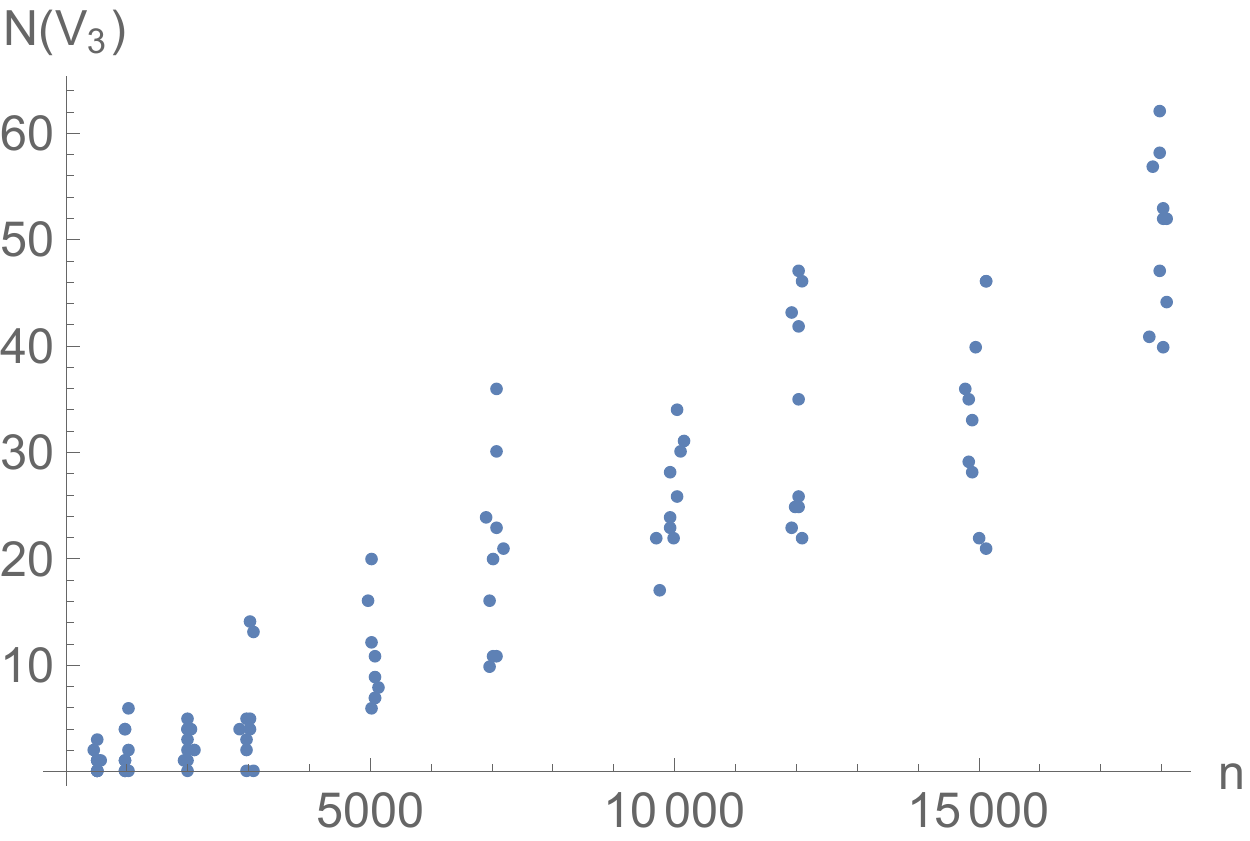}
  \caption{The abundance of $V_3$ as a function of $n$.}\label{v-fig.fig}
\end{figure}
  \begin{figure}[H]
\centering                                                          \includegraphics[height=5cm]{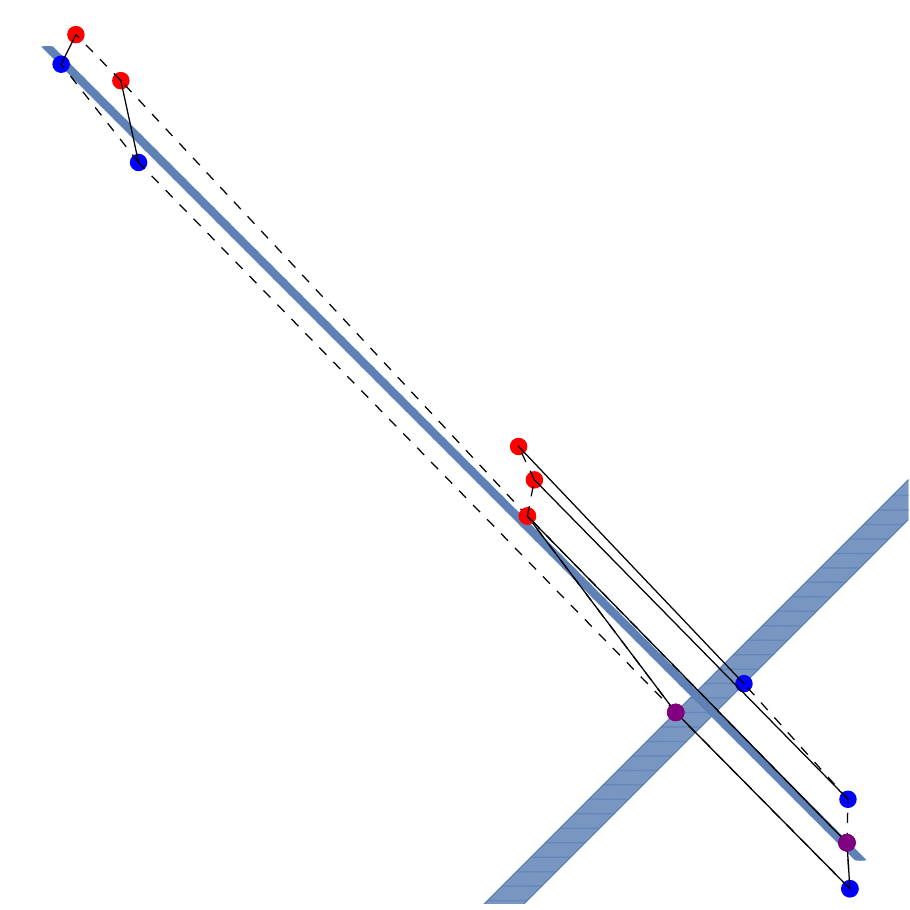} 
                
                \caption{An example of an $V_4$ subcausal set, using which we obtain an intrinsic characterisation of
                  the directions $\partial_{\mathrm u}, \partial_{\mathrm v}$. The purple points are the pairs
                  $(q_1,p_2)=(p_2',q_1')$. As before, the $p_i, p_i'$ are depicted in blue and the $q_i,q_i'$ in red.} \label{vex.fig}
 \end{figure}

Finally, our simulations support the conjecture  that there is a unique $L_k$ between  $h \dhor h'$. In all our simulations,
we find that there are no ``branching'' $L_k$ (except of the above $V_k$ sort)  {\it upto} the last $q_k$. In other
words, the only cases we find are ladders which are identical but for a single differing element, namely $q_k$,  
examples of which we show in Fig \ref{unique.fig}.
\begin{figure}[H]
  \centerline{\begin{tabular}{cc}
 \includegraphics[height=4cm]{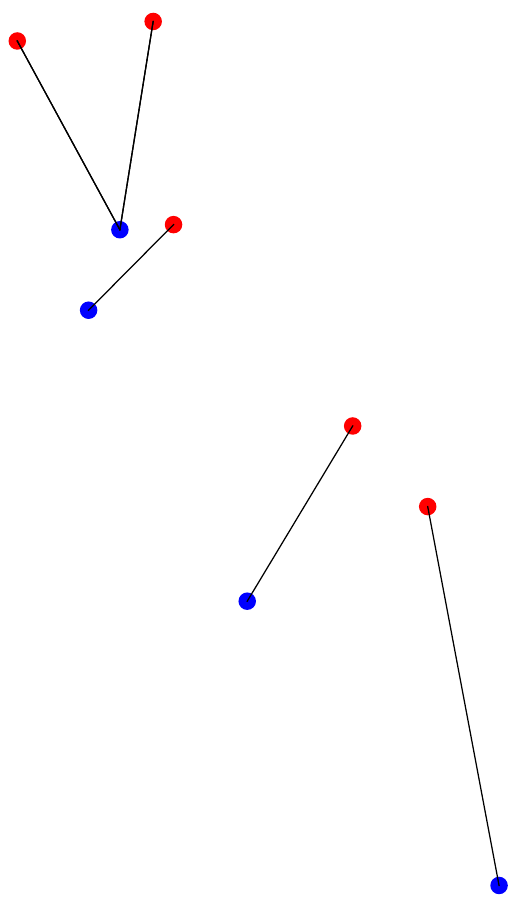} &  \hskip 3cm
                                                         \includegraphics[height=4cm]{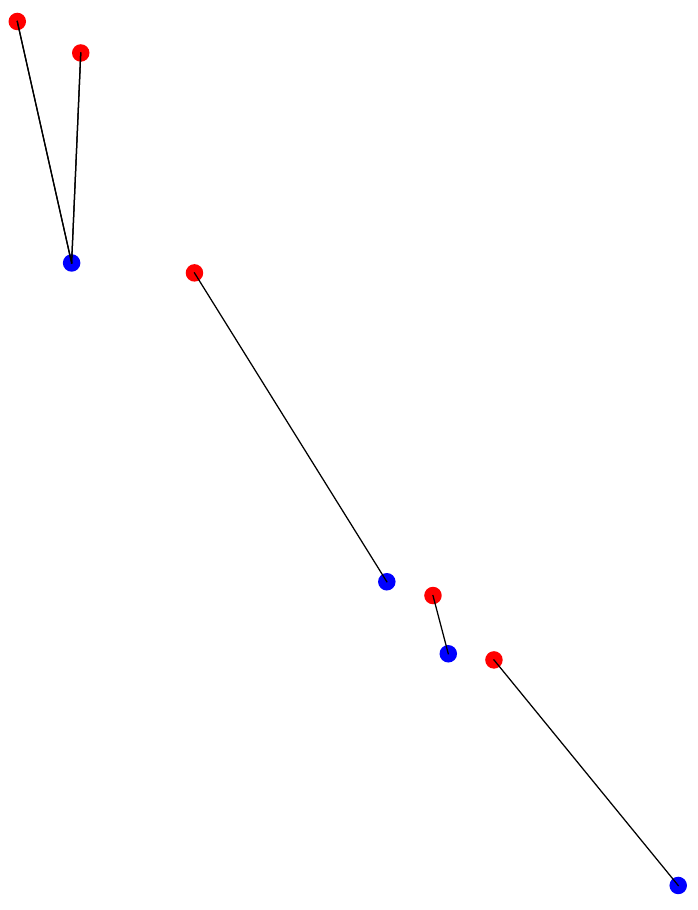}
                \\
                 \end{tabular}}
                \caption{Two examples of $L_4$, which are unique ``upto'' $q_4$. Given $h_1$ and $h_4$, there is a
                  unique $L_4$ from $h_1$ to $h_4$. However $h_4$ itself is not unique and so we have a second ladder $L_4'$
                from $h_1$ to $h_4'$. Such non-uniqueness however gets ironed out as one goes to higher $k$, as is the
                case with the $L_3$'s.  } \label{unique.fig}
                
\end{figure} 
Proposition 2.19 of Penrose's monograph \cite{penrose} states that:
\begin{proposition*}
(Penrose) If $\alpha$ is a null geodesic from $a$ to $b$ and $\beta$ is a null geodesic from $b$  to $c$, then either
$a\pprec b$ or else $\alpha \cup \beta$ constitutes a single null geodesic from $a$ to $c$.    
  \end{proposition*} 
 \noindent Our simulations lend support to the following  generalisation thus providing the strongest evidence that ladders
 are indeed the analogues of null  geodesics in causal sets that embed into $\mink^2$: 
\begin{conjecture*}
If $L_k, k>1$ is a ladder from $h$ to $h'$ and $L_{k'}', k'>1$ is a ladder from $h'$  to $h''$, where the $h,h',h''$ are
linked pairs of elements, then either $h \not \dhor h'' $ or else $L_k \cup L_{k'}'$ constitutes a single ladder
$L_{k+k'-1}$ from $h$ to $h''$.      
\end{conjecture*}

We have taken  the  first  steps towards a  geometric reconstruction of 
null geodesics in manifold like causal sets. The  construction presented above is limited to $\mink^2$ but because all spacetimes in $d=2$ are conformally related it is  likely that this construction will go
through for curved $d=2$ spacetimes.  The generalisation to higher dimensions is non-trivial and  currently being
investigated. Any definition of  null-ladders should be able to trap an entire $d$-dimensional pencil of 
null-geodesics  for it to be realised in the causal set via a Poisson process. Hence the $d=2$ null ribbon construction
cannot suffice in higher dimensions. A proposal currently being investigated is to use $(d-1)$ horizon molecules as the
rungs of the null ladder with  $p \link q^{(i)}, i=1, \ldots d-1$  forming  a  $(d-1)$``simplex'' whose $(d-1)$ face has
a space-like normal. Constructing the ladder molecule requires the $q^{(i)}$ from different rungs to be tied together
suitably in order to make the ladder as straight as possible. Numerical
investigations of such a ladder molecule construction in $d=3$ are underway.

Our work is numerical, more out of simplicity than necessity.  Integrals for calculating the probabilities of ladder
molecules are easily defined but non-trivial to calculate analytically, since several simultaneous conditions must be
met by each ladder element. While the   interval abundance calculations of 
\cite{intervals} give complex, but closed form expressions, preliminary work on ladder molecules suggests otherwise.
Numerical investigations of this causal set architecture will therefore remain a  robust tool in investigating
generalised ladder molecules in higher dimensions.  

\noindent {\bf Acknowledgements:} We thank  the organisers  of the conference SCRI21 (where one of the authors was a speaker) for their 
invitation to submit a contribution to  this collection. This work as well as many aspects of the causal set approach to quantum gravity are
strongly influenced by 
Roger Penrose's ideas on causal structure.

\noindent {\sl The datasets generated during and/or analysed during the current study are available from the corresponding author on reasonable request. } 

\bibliography{Refs}
\bibliographystyle{ieeetr}
\end{document}